\documentclass[twocolumn,aps]{revtex4}
\usepackage{amsmath}
\usepackage{graphicx}
\usepackage{dcolumn}
\usepackage{bm}

\begin{document}

\title{Photonic band structures of periodic arrays of pores in a metallic
host: tight-binding beyond the quasistatic approximation}


\author{Kwangmoo Kim$^{1,2}$ and D. Stroud$^1$}
\affiliation{$^1$Department of Physics, The Ohio State University, Columbus, Ohio 43210, USA \\
$^2$School of Physics, Korea Institute for Advanced Study, Seoul, 130-722, Republic of Korea}


\date{\today}

\begin{abstract}
We have calculated the photonic band structures of metallic inverse
opals and of periodic linear chains of spherical pores in a metallic host, below a 
plasma frequency $\omega_{\text{p}}$.  In both cases, we use a tight-bind\-ing
approximation, assuming a Drude dielectric
function for the metallic component, but without making the quasistatic
approximation.  The tight-binding modes are linear combinations of the 
single-cavity transverse magnetic (TM) modes.
For the inverse-opal structures, the lowest modes are analogous to those 
constructed from the three degenerate atomic $p$-states in fcc crystals.  
For the linear chains, in the limit of small spheres compared to
a wavelength, the results are the ``inverse'' of the dispersion
relation for metal spheres in an insulating host, as calculated
by Brongersma {\it et al.} [Phys.\ Rev.\ B \textbf{62}, R16356 (2000)].
Because the electromagnetic fields of these modes decay exponentially in the
metal, there are no radiative losses, in contrast to the case of arrays of 
metallic spheres in air.
We suggest that this tight-binding approach to photonic band
structures of such metallic inverse materials may be a useful approach
for studying photonic crystals containing metallic components, even beyond
the quasistatic approximation.

\end{abstract}


\maketitle


\section{\label{sec:level1}Introduction}

The photonic band structures of composite materials have been
studied extensively.  Such band structures are defined by the
relation between frequency $\omega$ and Bloch vector $\mathbf{k}$ in
media in which the dielectric constant is a periodic function of
position.  A major reason for such interest is the possibility of
producing photonic band gaps, i.e., frequency regions, extending
through all $\mathbf{k}$-space, where electromagnetic waves cannot
propagate through the medium.   Such media have many potentially
valuable applications, including possible use as filters and in
films with re\-jec\-tion-wave\-length tuning.\cite{xia}  In systems with a
complete photonic band gap, the spontaneous emission of atoms with
level splitting within the gap can be strongly suppressed.\cite{john}

Since light cannot travel through the photonic band gap materials (Bragg diffracted backwards), one of their
applications can be a complete control over wasteful spontaneous emission in unwanted directions when a
device, such as a laser, is embedded inside a 3D photonic crystal.\cite{yablonovitch} 2D photonic crystals can be used as optical
microcavities, microresonators,\cite{scherer} waveguides,\cite{mekis}
lasers, \cite{painter} or fibers\cite{benabid} while 1D photonic crystals can be used as Bragg gratings or optical
switches.\cite{cao}

The photonic band structure of a range of materials has been studied
using a plane wave expansion method.  Typically, the method
converges easily when the dielectric function is everywhere real,
but more slowly, or not at all, when the dielectric function has a
negative real part, as occurs when one component is metallic.  For
example, McGurn {\it et al.}\cite{mcgurn} used this method to
calculate the photonic band structure of a square lattice of metal
cylinders in two dimensions (2D) and of an fcc lattice of metal
spheres embedded in vacuum in 3D.   They found that that method
converged well when the filling fraction $f$ (i.e., volume fraction
of metal spheres or cylinders) satisfied $f\leq 0.1\%$.

Kuzmiak {\it et al.}\cite{kuzmiak1} used the same method to
calculate the photonic band structures for 2D metal cylinders in a
square or triangular lattice in vacuum.   For low $f$ and $\omega >
\omega_{\text{p}}$, the calculated photonic band structures are just slightly
perturbed versions of the dispersion curves for electromagnetic
waves in vacuum.   However, for $\omega < \omega_{\text{p}}$ and
$\mathbf{H}$-po\-lar\-ized waves (magnetic field $\mathbf{H}$ parallel to the
cylinders), they obtained many nearly flat bands for
$\omega<\omega_{\text{p}}$; these bands were found to converge
very slowly with increasing numbers of plane waves.  They later
extended this work to systems with dissipation.\cite{kuzmiak2}  To
describe dispersive and absorptive materials, they used a complex,
po\-si\-tion-de\-pend\-ent form of dielectric function.  They also
introduced a standard linearization technique to solve the resulting
nonlinear eigenvalue problem.


Zabel {\it et al.}\cite{zabel} extended the plane wave method to
treat periodic composites with anisotropic dielectric functions. In
particular, they studied the photonic band structures of a periodic
array of anisotropic dielectric spheres embedded in air.  They found
that the anisotropy split degenerate bands, and narrowed or even
closed the band gaps.  Much further work on anisotropic photonic
materials has been carried out since this paper (see, e.g., Ref.\ \cite{john}).

A different type of periodic metal-insulator composite is a periodic arrangement
of metallic spheres in an insulating host.  Brongersma {\it et al.}\cite{brongersma}
studied the dispersion relation for coupled plasmon modes in such a linear chain of
equally spaced metal nanoparticles, using a near-field electromagnetic (EM) interaction
between the particles in the dipole limit. They also studied the transport of EM energy
around the corners and through tee junctions of the nanoparticle chain-array.

Park and Stroud\cite{park} also studied the surface-plasmon dispersion relations for
a chain of metallic nanoparticles in an isotropic medium. They used
a generalized tight-binding calculations, including all multipoles.  This approach
is more exact than the previous point-dipole calculation,\cite{brongersma}
in a quasistatic limit, but still leaves out non-quasistatic effects associated
with radiative damping (i.e., effects associated with the non-vanishing of
$\mathbf{\nabla} \times \mathbf{E}$, where $\mathbf{E}$ is the electric field.
They calculated the lowest bands as well as many
higher bands and compared their results with those in Ref.\ \cite{brongersma}.

Weber and Ford\cite{weber} have shown that all calculations within the quasistatic
approximation omit important interactions between transverse plasmon waves and
free photon modes, even if the interparticle separation is small compared to the
wavelength of light.  Thus, most quasistatic calculations need to have certain
corrections included at particular values of the wave vector.

Recently, Gaillot {\it et al.}\cite{gaillot} have studied the
photonic band structures of another type of structure, a so-called
inverse opal structure.    This structure is an fcc lattice of void
spheres in a host of another material.   Such a structure can be
prepared, e.g., starting from an opal structure made of spheres of
a convenient substance, infiltrating it with another material, then
dissolving away the spheres.
In the work of Ref.\ \cite{gaillot}, the photonic band
structure of Si inverse opal was calculated as a function of the
infiltrated volume fraction $f$ of air voids using three-di\-men\-sion\-al
finite difference time domain (3D FDTD) method.  It was found that
for certain values of $f$, a complete band gap opens up between the
eighth and ninth bands.

In the present work, first we study the photonic band structure of an
inverse opal structure, such as that investigated in Ref.\ \cite{gaillot},
but instead of dielectric materials such as Si, we
consider metals as the infiltrated materials.  Thus, the material we
study is also the inverse of the fcc array of metal spheres studied
by McGurn {\it et al.}\cite{mcgurn}  Such metallic inverse opal
structures have recently become of great interest, because it has
been found that Pb inverse opals exhibit superconductivity.\cite{aliev}
These workers have studied the response of these
materials to an applied magnetic field, and have found a highly
non-mon\-o\-ton\-ic fractional flux penetration into the Pb spheres as a
function of the applied field.

As a second example, we study the photonic band structure
of a linear chain of nanopores in a metallic medium. This is an inverse
structure of a linear chain of metallic nanospheres, of which the dispersion
relation is given in Ref.\ \cite{brongersma}. As anticipated, we get a kind of
``inverse image'' of the dispersion relation found by Ref.\ \cite{brongersma}
in our system.

For both types of structures, our primary method
for studying the photonic band structures below the plasma
frequency $\omega_{\text{p}}$ is a tight-bind\-ing approximation which is
valid even in the non-quasistatic regime.  Because the analogs of the tight-binding
atomic states decay exponentially in the metallic host medium,
the resulting tight-binding waves do not lose energy radiatively, as do the
corresponding waves along one-dimensional chains of metallic nanoparticles in
air.  Furthermore, because the modes are expanded in ``atomic'' states rather
than plane waves, there is no convergence problem as there can be in the plane
wave case.


The remainder of this paper is organized as follows. In Section
\ref{sec:level2}, we first present the formalism for calculating
the transverse magnetic (TM) and transverse electric (TE) modes of a
single spherical cavity in a metallic host.  We then describe the
method for calculating the photonic band structures of metallic
inverse opals and of linear chains of nanopores in a metallic host,
using a simple tight-bind\-ing approach for $\omega < \omega_{\text{p}}$.
In Section \ref{sec:level3}, we give the numerical results for the TM and
TE modes of a single cavity and those of the tight-bind\-ing method for
the metal inverse opals and the linear chain of nanopores.
Section \ref{sec:level4} presents a summary and discussion.

\section{\label{sec:level2}Formalism}

In this section, we present a summary of the equations determining
the band structure of a photonic crystal containing a metallic
component with Drude dielectric function $\epsilon(\omega) = 1 -
\omega_{\text{p}}^2/\omega^2$ and an insulating component of dielectric
constant unity.  The insulating component is assumed to be present in
the form of identical spherical cavities of radius $R$.
We first write down the equations for the TM and TE
modes of a spherical cavity in a Drude metal. Then, we present
a tight-bind\-ing method for $\omega < \omega_{\text{p}}$.

\subsection{\label{subsec:level21}Spherical Cavity}

As a preliminary to calculating the photonic band structure, we
first discuss the modes of a single spherical cavity in a Drude metal
host. 
We begin with the TM modes of the cavity, then the TE modes.

\subsubsection{\label{subsubsec:level211}TM Modes}

It is convenient to describe the modes of the embedded cavity in terms
of the $\mathbf{B}$ field.
To that end, we combine the two homogeneous Maxwell equations
\begin{eqnarray}
\mathbf{\nabla}\times \mathbf{E} & = & \frac{i\omega}{c}\mathbf{B}, \label{eq:curle} \\
\mathbf{\nabla}\times \mathbf{B} & = & -\frac{i\omega}{c}\epsilon\mathbf{E},
\label{eq:curlb}
\end{eqnarray}
to obtain a single equation for $\mathbf{B}$:
\begin{equation}
\mathbf{\nabla}\times
\left[\left\{\frac{1}{1-\frac{\omega_{\text{p}}^2}{\omega^2}}
\theta(\mathbf{x})+1-\theta(\mathbf{x})\right\}\mathbf{\nabla}\times \mathbf{
B}\right]=\frac{\omega^2} {c^2}\mathbf{B}.
\end{equation}
Here, we have expressed the position- and frequency-dependent dielectric function $\epsilon(\mathbf{x}, \omega)$  as
$1/\epsilon(\mathbf{x}, \omega) = \theta(\mathbf{x})/(1
- \omega_{\text{p}}^2/\omega^2) + 1 - \theta(\mathbf{x})$, where the step function
$\theta(\mathbf{x})=1$ inside the metallic region and $\theta(\mathbf{x}) = 0$ elsewhere.
Multiplying this equation by $\omega^2 - \omega_{\text{p}}^2$ and simplifying,
we obtain
\begin{equation}
[\omega^2-\omega_{\text{p}}^2(1-\theta(\mathbf{x}))]\mathbf{\nabla}\times (\mathbf{\nabla}
\times \mathbf{B})=\frac{\omega^2}{c^2}(\omega^2-\omega_{\text{p}}^2)\mathbf{B}.
\label{eq:sc}
\end{equation}
Thus, inside the spherical void, we have
\begin{equation}
\mathbf{\nabla}\times (\mathbf{\nabla} \times \mathbf{B})=\frac{\omega^2}{c^2}\mathbf{B},
\label{eq:void}
\end{equation}
while inside the metal,
\begin{equation}
\mathbf{\nabla}\times (\mathbf{\nabla} \times \mathbf{B})
=\frac{\omega^2-\omega_{\text{p}}^2}{c^2}\mathbf{B}.
\label{eq:metal}
\end{equation}

For a spherical void within a metallic host, it is convenient to solve these equations in spherical coordinates.   It is readily
found that, for the TM modes,
the non-vanishing components of the solutions for $\mathbf{B}$ and
$\mathbf{E}$ for Eq.\ (\ref{eq:void}) and Eq.\ (\ref{eq:curlb}) are\cite{jackson}
\begin{eqnarray}
& & B_{\phi,\text{in}}(r,\theta)=\frac{u_\ell(r)}{r}P_{\ell}^{1}(\cos\theta), \nonumber \\
& & E_{r,\text{in}}=-\frac{ic}{\omega r} \ell(\ell+1)\frac{u_{\ell}(r)}{r}
P_{\ell}(\cos \theta), \nonumber \\
& & E_{\theta,\text{in}}=-\frac{ic}{\omega r}\frac{\partial u_{\ell}(r)}
{\partial r}P_{\ell}^{1}(\cos \theta), \nonumber \\
& & u_{\ell}(r)=r[A_{\ell}j_{\ell}(kr)+B_{\ell}n_{\ell}(kr)],
\label{eq:ul}
\end{eqnarray}
where $k=\omega/c$, $j_{\ell}$ and $n_{\ell}$ are spherical
Bessel functions, and the subscripts $\phi$, $r$, and $\theta$
denote components of the corresponding fields in spherical
coordinates.

Likewise, the solutions of Eqs.\ (\ref{eq:metal}) and (\ref{eq:curlb}) within the metal will be
\begin{eqnarray}
& & B_{\phi,\text{out}}(r,\theta)=\frac{v_\ell(r)}{r}P_{\ell}^{1}(\cos\theta), \nonumber \\
& & E_{r,\text{out}}=-\frac{ic}{\omega r \left(1-\frac{\omega_{\text{p}}^2}{\omega^2}\right)}
\ell(\ell+1)\frac{v_{\ell}(r)}{r} P_{\ell}(\cos \theta), \nonumber \\
& & E_{\theta,\text{out}}=-\frac{ic}{\omega r \left(1-\frac{\omega_{\text{p}}^2}{\omega^2}\right)}
\frac{\partial v_{\ell}(r)}{\partial r}P_{\ell}^{1}(\cos \theta), \nonumber \\
& & v_{\ell}(r)=r[C_{\ell}j_{\ell}(k^\prime r)+D_{\ell}n_{\ell}(k^\prime r)],
\label{eq:vl}
\end{eqnarray}
where $k^\prime=\sqrt{\omega^2-\omega_{\text{p}}^2}/c$.

The requirements that the normal displacement $D$ and tangential $E$ should be
continuous at $R$ gives the two conditions
\begin{equation}
u_{\ell}(R)=v_{\ell}(R)
\label{eq:ne1}
\end{equation}
and
\begin{equation}
\left. \frac{\partial u_{\ell}(r)}{\partial r}\right|_{r=R}
=\frac{1}{\left(1-\frac{\omega_{\text{p}}^2}{\omega^2}\right)}
\left. \frac{\partial v_{\ell}(r)}{\partial
r}\right|_{r=R}=\frac{k^2}{k^{\prime 2}} \left. \frac{\partial
v_{\ell}(r)}{\partial r}\right|_{r=R},\label{eq:te1}
\end{equation}
where $R$ is the radius of the spherical cavity.  Since the fields
at the center of the void sphere must be finite, we also have
\begin{equation}
u_{\ell}(r)=rj_{\ell}(kr),
\end{equation}
where we have normalized the solution so that the coefficient
$A_{\ell} = 1$.  From Eq.\ (\ref{eq:ne1}) we have
\begin{equation}
j_{\ell}(kR)=C_{\ell}j_{\ell}(k^\prime R)+D_{\ell}n_{\ell}(k^\prime R),
\label{eq:ne3}
\end{equation}
whereas, from Eq.\ (\ref{eq:te1}), we get
\begin{eqnarray}
& & k^{\prime 2}\left[j_{\ell}(kR)+kRj_{\ell}^\prime(kR)\right] \nonumber \\
& = & k^2\left[C_{\ell}j_{\ell}(k^\prime R)+D_{\ell}n_{\ell}(k^\prime R) + k^\prime R\left\{C_{\ell}
j_{\ell}^\prime(k^\prime R) \right.\right. \nonumber \\
& & + \left.\left. D_{\ell}n_{\ell}^\prime(k^\prime R)\right\}\right].
\label{eq:te3}
\end{eqnarray}
The coefficients $C_{\ell}$ and $D_{\ell}$ can then be determined from these
boundary conditions.

The results for $\omega < \omega_{\text{p}}$, can be obtained by
making the substitution $k^\prime \rightarrow ik^\prime$, with
$k^\prime$ real.  In this case, the radial component of Eq.\
(\ref{eq:metal}) takes the form
\begin{equation}
\frac{\partial^2 u_{\ell}}{\partial r^2}-\left[\frac{1}{c^2}(\omega_{\text{p}}^2
-\omega^2)+\frac{\ell(\ell+1)}{r^2}\right]u_{\ell}=0,
\label{eq:metalr}
\end{equation}
which is the modified Bessel equation. The solutions of Eq.\
(\ref{eq:metalr}) are the modified spherical Bessel functions
$i_{\ell}$ and $k_{\ell}$ (note that this $k_{\ell}$ is different
from the wave vectors $k$ and $k^\prime$).  In this case, Eq.\
(\ref{eq:vl}) becomes
\begin{equation}
v_{\ell}(r)=r[C_{\ell}i_{\ell}(k^\prime r)+D_{\ell}k_{\ell}(k^\prime r)],
\label{eq:vl1}
\end{equation}
where $k^\prime=\sqrt{\omega_{\text{p}}^2-\omega^2}/c$.  In
addition, Eqs.\ (\ref{eq:te1}), (\ref{eq:ne3}), and (\ref{eq:te3})
are transformed, respectively, into
\begin{equation}
\left. \frac{\partial u_{\ell}(r)}{\partial r}\right|_{r=R}
=-\frac{k^2}{k^{\prime 2}}
\left. \frac{\partial v_{\ell}(r)}{\partial r}\right|_{r=R},
\label{eq:te4}
\end{equation}
\begin{equation}
j_{\ell}(kR)=C_{\ell}i_{\ell}(k^\prime R)+D_{\ell}k_{\ell}(k^\prime R),
\label{eq:ne4}
\end{equation}
and
\begin{eqnarray}
& & k^{\prime 2}\left[j_{\ell}(kR)+kRj_{\ell}^\prime(kR)\right]   =  -
k^2\left[C_{\ell}i_{\ell}(k^\prime R) \right. \nonumber \\
& & \left. +D_{\ell}k_{\ell}(k^\prime
R)+k^\prime R\left\{C_{\ell} i_{\ell}^\prime(k^\prime R) +
D_{\ell}k_{\ell}^\prime(k^\prime R)\right\}\right]. \label{eq:te5}
\end{eqnarray}

It is of interest to consider the specific case of a spherical
cavity in an infinite medium.
In this case, $C_{\ell}=0$ because $i_{\ell}(x)$ diverges at large $x$.
As a result, Eqs.\ (\ref{eq:vl1}), (\ref{eq:ne4}), and
(\ref{eq:te5}) become, respectively,
\begin{equation}
v_{\ell}(r)=r D_{\ell}k_{\ell}(k^\prime r),
\label{eq:vl2}
\end{equation}
\begin{equation}
j_{\ell}(kR)=D_{\ell}k_{\ell}(k^\prime R),
\label{eq:ne5}
\end{equation}
and
\begin{eqnarray}
& & k^{\prime 2}\left[j_{\ell}(kR)+kRj_{\ell}^\prime(kR)\right]
= - k^2 D_{\ell}\left[k_{\ell}(k^\prime R) \right. \nonumber \\
& & \left. +k^\prime R k_{\ell}^\prime(k^\prime R)\right].
\label{eq:te6}
\end{eqnarray}
From Eq.\ (\ref{eq:ne5}) we have
$D_{\ell}=j_{\ell}(kR)/k_{\ell}(k^\prime R)$, and hence Eq.\ (\ref{eq:te6}) becomes
\begin{eqnarray}
& & k^{\prime 2}\left[j_{\ell}(kR)+kRj_{\ell}^\prime(kR)\right]
= - k^2\frac{j_{\ell}(kR)}{k_{\ell}(k^\prime R)}\left[k_{\ell}(k^\prime R)\right. \nonumber \\
& & \left. +k^\prime R k_{\ell}^\prime(k^\prime R)\right].
\label{eq:ws4}
\end{eqnarray}

We can readily obtain the asymptotic forms of the solutions when
$kR\ll 1$ and $k^\prime R \ll 1$.  In this case $j_{\ell}(kR)$ and
$k_{\ell}(k^\prime R)$ have the asymptotic forms
$j_{\ell}(kR)\approx (kR)^{\ell}/(2\ell+1)!!$ and
$k_{\ell}(k^\prime R)\approx \sqrt{\pi}\Gamma(\ell+1/2)
2^{\ell-1}/(k^\prime R)^{\ell+1}$.  In this limit, Eq.\ (\ref{eq:ws4}),
after some algebra, reduces to simply
\begin{equation}
k^{\prime 2}(\ell+1)=k^2 \ell.
\label{eq:asympt1}
\end{equation}
Since $k^\prime=\sqrt{\omega_{\text{p}}^2-\omega^2}/c$, Eq.\
(\ref{eq:asympt1}) is equivalent to
\begin{equation}
\omega^2=\frac{\ell+1}{2\ell+1}\omega_{\text{p}}^2.
\end{equation}
The largest value, $\omega = \sqrt{2/3}\omega_{\text{p}}$, occurs at $\ell=1$
and the limiting value for large $\ell$ is $\omega=\omega_{\text{p}}/\sqrt{2}$.

\subsubsection{\label{subsubsec:level212}TE Modes}

For the TE mode, inside the spherical void, we have
\begin{equation}
\mathbf{\nabla}\times (\mathbf{\nabla} \times \mathbf{E})=\frac{\omega^2}{c^2}\mathbf{E},
\label{eq:voidte}
\end{equation}
whereas inside the metal, we have
\begin{equation}
\mathbf{\nabla}\times (\mathbf{\nabla} \times \mathbf{E})
=\frac{\omega^2-\omega_{\text{p}}^2}{c^2}\mathbf{E}.
\label{eq:metalte}
\end{equation}

We now use these equations to calculate $E_{\phi}$, $B_{r}$, and
$B_{\theta}$. From Eq.\ (\ref{eq:voidte}) we get
\begin{eqnarray}
& & E_{\phi,\text{in}}(r,\theta)=\frac{u_\ell(r)}{r}P_{\ell}^{1}(\cos\theta), \nonumber \\
& & B_{r,\text{in}}=\frac{ic}{\omega r} \ell(\ell+1)\frac{u_{\ell}(r)}{r}
P_{\ell}(\cos \theta), \nonumber \\
& & B_{\theta,\text{in}}=\frac{ic}{\omega r}\frac{\partial u_{\ell}(r)}
{\partial r}P_{\ell}^{1}(\cos \theta), \nonumber \\
& & u_{\ell}(r)=r[A_{\ell}j_{\ell}(kr)+B_{\ell}n_{\ell}(kr)],
\label{eq:ul3}
\end{eqnarray}
where $k=\omega/c$.
The solutions of Eq.\ (\ref{eq:metalte}) are
\begin{eqnarray}
& & E_{\phi,\text{out}}(r,\theta)=\frac{v_\ell(r)}{r}P_{\ell}^{1}(\cos\theta), \nonumber \\
& & B_{r,\text{out}}=\frac{ic}{\omega r}
\ell(\ell+1)\frac{v_{\ell}(r)}{r} P_{\ell}(\cos \theta), \nonumber \\
& & B_{\theta,\text{out}}=\frac{ic}{\omega r}
\frac{\partial v_{\ell}(r)}{\partial r}P_{\ell}^{1}(\cos \theta), \nonumber \\
& & v_{\ell}(r)=r[C_{\ell}j_{\ell}(k^\prime r)+D_{\ell}n_{\ell}(k^\prime r)],
\label{eq:vl3}
\end{eqnarray}
where $k^\prime=\sqrt{\omega^2-\omega_{\text{p}}^2}/c$.

Since normal $B$ and tangential $H$ should be continuous on the
boundaries, we obtain the conditions
\begin{equation}
u_{\ell}(R)=v_{\ell}(R),
\label{eq:nb1}
\end{equation}
as in the TM case, and
\begin{equation}
\left. \frac{\partial u_{\ell}(r)}{\partial r}\right|_{r=R} =\left.
\frac{\partial v_{\ell}(r)}{\partial r}\right|_{r=R}. \label{eq:th1}
\end{equation}
Since the fields must be finite at the center of the void sphere, we
can choose
\begin{equation}
u_{\ell}(r)=rj_{\ell}(kr),
\end{equation}
we also take the coefficient $A_{\ell} = 1$.  From Eq.\ (\ref{eq:nb1})
we have
\begin{equation}
j_{\ell}(kR)=C_{\ell}j_{\ell}(k^\prime R)+D_{\ell}n_{\ell}(k^\prime R),
\label{eq:nb2}
\end{equation}
while, from Eq.\ (\ref{eq:th1}), we get
\begin{eqnarray}
& & j_{\ell}(kR)+kRj_{\ell}^\prime(kR) \nonumber \\
& = & C_{\ell}j_{\ell}(k^\prime R)+D_{\ell}n_{\ell}(k^\prime R)+k^\prime R\left[C_{\ell}
j_{\ell}^\prime(k^\prime R) \right. \nonumber \\
& & \left. + D_{\ell}n_{\ell}^\prime(k^\prime R)\right].
\label{eq:th2}
\end{eqnarray}

The corresponding equation for $\omega < \omega_{\text{p}}$, can again be
obtained by the transformation $k^\prime \rightarrow ik^\prime$. The
TE modes for $\omega < \omega_{\text{p}}$ using the modified
spherical Bessel functions $i_{\ell}(x)$ and $k_{\ell}(x)$ satisfy
Eqs.\ (\ref{eq:metalr}), (\ref{eq:vl1}), (\ref{eq:th1}), and (\ref{eq:ne4}).
The only changes are in Eq.\ (\ref{eq:te5}),
which becomes
\begin{eqnarray}
& & j_{\ell}(kR)+kRj_{\ell}^\prime(kR)
 =  C_{\ell}i_{\ell}(k^\prime R)+D_{\ell}k_{\ell}(k^\prime R) \nonumber \\
& & +k^\prime R\left\{C_{\ell}
i_{\ell}^\prime(k^\prime R) + D_{\ell}k_{\ell}^\prime(k^\prime R)\right\}.
\label{eq:th3}
\end{eqnarray}
In an infinite medium, these conditions
become, from Eq.\
(\ref{eq:ne5}),
\begin{equation}
j_{\ell}(kR)+kRj_{\ell}^\prime(kR) =
\frac{j_{\ell}(kR)}{k_{\ell}(k^\prime R)}\left[k_{\ell}(k^\prime R)
+k^\prime R k_{\ell}^\prime(k^\prime R)\right]. \label{eq:wsteinfty}
\end{equation}

If we consider the asymptotic forms of the solutions when
$kR\ll 1$ and $k^\prime R \ll 1$ as we did for the TM modes, Eq.\ (\ref{eq:wsteinfty})
simplifies to
\begin{equation}
\ell+1=-\ell,
\label{eq:asympt2}
\end{equation}
which gives $\ell=-1/2$.  Since $\ell$ must be a positive integer,
we see that there are no eigenvalues for TE modes in
the limit $kR \ll 1$ and $k^\prime R \ll 1$.

\subsection{\label{subsec:level22}Tight-Bind\-ing Approach to Modes
for $\omega < \omega_{\text{p}}$}

We now turn from describing the single-cavity modes to a discussion of
the band structure for a periodic array of such cavities.  In conventional
periodic solids, the tight-bind\-ing method is very useful in treating narrow
bands.  In what follows, we try to suggest an analogous
tight-bind\-ing approach for the lowest set of TM modes in a periodic
lattice of spherical cavities in a metallic host, in the frequency range
$\omega < \omega_{\text{p}}$.  We apply the resulting method, first, to an fcc lattice
of pores, and then to a linear chain of spherical pores in a metallic host.


Even though these are TM modes, it is convenient to describe them now in terms
of their electric fields.  We denote the electric field of the $\lambda$th mode
by $\mathbf{E}_\lambda(\mathbf{x})$.   This field satisfies
\begin{equation}
\mathbf{\nabla}\times(\mathbf{\nabla}
\times \mathbf{E}_\lambda(\mathbf{x})) + \frac{\omega_{\text{p}}^2\theta(\mathbf{
x})}{c^2}\mathbf{E}_\lambda(\mathbf{x}) \equiv \mathcal{O}\mathbf{E}_\lambda(\mathbf{x}) 
= \frac{\omega_\lambda^2}{c^2}\mathbf{E}_\lambda(\mathbf{x}),
\label{eq:eigenv}
\end{equation}
where $\mathcal{O} = \mathbf{\nabla}\times (\mathbf{\nabla} \times) + (\omega_{\text{p}}^2/c^2)\theta(\mathbf{x})$ 
is the ``Hamiltonian'' of this system.  Since $\mathcal{O}$ is a Hermitian operator,
the eigenstates corresponding to unequal eigenvalues $\omega_\lambda^2/c^2$ and $\omega_\mu^2/c^2$
are orthogonal and may be chosen to be orthonormal. (The orthogonality may also be proved
directly by integration by parts.)  The orthonormality relation is
\begin{equation}
\int\mathbf{E}_\lambda^*(\mathbf{x})\cdot\mathbf{E}_\mu(\mathbf{x})d\mathbf{x} =
\delta_{\lambda,\mu}.
\label{eq:orthon}
\end{equation}
Since $\mathbf{E}_\lambda(\mathbf{x})$ is real for $\omega < \omega_{\text{p}}$, the complex conjugation 
is, in fact, unnecessary.

In Sec.\ \ref{subsubsec:level211}, our paper already gives the equations determining the electric 
and magnetic fields of isolated TM
modes for a spherical cavity.  The lowest set corresponds to $\ell = 1$, and there
should be three of these.  For a spherical cavity, all three
are degenerate, i.e., all three have the same eigenfrequencies.
Even though
the three modes have equal frequencies, one can always choose an orthonormal
set, with electric fields $\mathbf{E}_1$, $\mathbf{E}_2$, and $\mathbf{E}_3$ satisfying
the orthonormality relation in Eq.\ (\ref{eq:orthon}).

In order to obtain the tight-binding band structure built from these three modes, 
we need to calculate matrix elements of the form
\begin{equation}
M_{\alpha,\beta}(\mathbf{R}) = \int \mathbf{E}_{\alpha}^{*}(\mathbf{x})\cdot 
\mathcal{O}\mathbf{E}_\beta(\mathbf{x} - \mathbf{R})d\mathbf{x}, 
\label{eq:mab}
\end{equation}
corresponding to two single-cavity modes associated with different cavities
centered at the origin and at $\mathbf{R}$.  Here, $\mathcal{O}$ is the
``Hamiltonian'' of the system as defined implicitly in Eq.\ (\ref{eq:eigenv}).

Next, we introduce normalized Bloch states associated with
the three $\ell = 1$ sin\-gle-cav\-i\-ty modes.  In order to do this,
we first make the usual tight-bind\-ing assumption that the
``atomic'' states corresponding to different cavities are
orthogonal:
\begin{equation}
\int \mathbf{E}_\lambda^*(\mathbf{x} - \mathbf{R})\cdot\mathbf{E}_\mu(\mathbf{x} -
\mathbf{R}^\prime)d\mathbf{x} = \delta_{\lambda,\mu}\delta_{\mathbf{R},\mathbf{R}^\prime}.
\label{eq:orthog}
\end{equation}
This orthogonality of states on different cavities is reasonable since
the fields fall off exponentially with separation.

The orthonormal Bloch states then take the form
\begin{equation}
\mathbf{E}_{\mathbf{k},\lambda}(\mathbf{x}) = N^{-1/2}\sum_\mathbf{R}
e^{i\mathbf{k}\cdot\mathbf{R}}\mathbf{E}_\lambda(\mathbf{x}-\mathbf{R}),
\label{eq:nbs}
\end{equation}
where $\mathbf{k}$ is a Bloch vector, and the $\mathbf{R}$'s are the Bravais
lattice vectors.  In writing Eq.\ (\ref{eq:nbs}), we have assumed that there are
$N$ identical spherical cavities, and that the Bloch states satisfy the usual
periodic boundary conditions of Born-von Karman type.
We also introduce the elements of the ``Hamiltonian'' matrix
\begin{equation}
M_{\lambda,\mu}(\mathbf{k}) = \sum_\mathbf{R}e^{i\mathbf{k}
\cdot\mathbf{R}}M_{\lambda,\mu}(\mathbf{R}).
\end{equation}

We can then obtain the frequencies $\omega(\mathbf{k})$ by diagonalizing a $3 \times 3$ matrix as follows:
\begin{equation}
det\left|M_{\lambda,\mu}(\mathbf{k}) - \left(\frac{\omega^2(\mathbf{k})}{c^2}
-\frac{\omega_{\text{at}}^2}{c^2}\right)\delta_{\lambda,\mu}\right| = 0,
\end{equation}
where $\omega_{\text{at}}$ is the eigenvalue of a sin\-gle-cav\-i\-ty mode.
The solutions to these equations give the three $p$-bands for a
periodic lattice of cavities in a metallic host.  This procedure is analogous
to that used in the well-known procedure for obtaining tight-binding bands from
three degenerate $p$-bands in the electronic structure of conventional solids
(see, for example, Ref.\ \cite{ashcroft}).

We briefly comment on the connection between this approach and that
used by earlier workers.\cite{brongersma,park}  In this work, the
authors treat wave propagation along a chain of metallic
nanoparticles.  They use the tight-bind\-ing approximation, as we do,
but in the quasistatic approximation in which one assumes that
$\mathbf{\nabla}\times \mathbf{E} = 0$.  This approximation is reasonable when
both the particle radii and the interparticle separations are small
compared to a wavelength, but is not accurate in other
circumstances.  Furthermore, even in the small-par\-ti\-cle and
small-sep\-a\-ra\-tion regime, this approximation still fails to account
for the radiation which occurs at certain wave numbers and
frequencies.  The present approach would generalize this
tight-bind\-ing method to (a) three dimensions as well as one; (b)
pore modes instead of small particle modes; and most importantly (c)
larger pores and larger interparticle separations, via extension beyond
the quasistatic approximation.

Next, we discuss the numerical evaluation of the required matrix
elements, Eq.\ (\ref{eq:mab}).  The relevant electric fields are
given in this paper, but in spherical coordinates.  It is not
difficult to convert these into Cartesian coordinates.  The operator
$\mathcal{O}$ is just a little trickier.  We first note that $\mathcal{O}
= \mathcal{O}_\mathbf{R} + \mathcal{O}^\prime$, where $\mathcal{O}_\mathbf{R}$ is the
sin\-gle-cav\-i\-ty operator: 
$\mathcal{O}_\mathbf{R} = \mathbf{\nabla}\times (\mathbf{\nabla}\times$) if $\mathbf{x}$ is inside the $\mathbf{R}$th
cavity and $ \mathcal{O}_\mathbf{R} = \mathbf{\nabla} \times (\mathbf{\nabla} \times) + \omega_{\text{p}}^2/c^2$ 
otherwise.  Now we also have
\begin{equation}
\mathcal{O}_\mathbf{R}\mathbf{E}_\beta(\mathbf{x} - \mathbf{R}^\prime) =
\frac{\omega_{\text{at}}^2}{c^2}\mathbf{E}_\beta(\mathbf{x} - \mathbf{R}^\prime),
\end{equation}
since $\mathbf{E}_\beta$ is an eigenstate of $\mathcal{O}_\mathbf{R}$ with
an eigenvalue $\omega_{\text{at}}^2/c^2$.

But since we are assuming that the overlap integral between ``atomic'' electric 
field states centered on different sites vanishes, the term involving
$\mathcal{O}_\mathbf{R}$
does not contribute to the matrix element $M_{\alpha,\beta}$, which is therefore
just given by
\begin{equation}
M_{\alpha,\beta}(\mathbf{R}) = \int \mathbf{E}_\alpha(\mathbf{x})\cdot 
\mathcal{O}^\prime\mathbf{E}_\beta(\mathbf{x} - \mathbf{R})d\mathbf{x}.
\end{equation}
We can also write
\begin{equation}
\mathcal{O}^\prime = \frac{\omega_{\text{p}}^2}{c^2}\sum_{\mathbf{R}^\prime}
\theta_{\mathbf{R}^\prime}(\mathbf{x}), \label{eq:calop}
\end{equation}
where
\begin{equation}
\theta_\mathbf{R^\prime} = \theta(\mathbf{x} - \mathbf{R}^\prime),
\end{equation}
is a step function which is unity inside the cavity centered at
$\mathbf{R}^\prime$ and is zero otherwise.

A reasonable approximation to Eq.\ (\ref{eq:calop}) might be to
include just $\mathbf{R}^\prime = 0$.  In this case, we finally will
get
\begin{equation}
M_{\alpha,\beta}(\mathbf{R}) \sim \frac{\omega_{\text{p}}^2}{c^2}\int 
\mathbf{E}_\alpha(\mathbf{x})\cdot\mathbf{E}_\beta(\mathbf{x}-\mathbf{R})d\mathbf{x},
\end{equation}
where the integral runs just over the cavity centered at the origin.
As a further approximation, we can just replace $\mathbf{E}_\beta(\mathbf{x}
-\mathbf{R})$ by the value of this function at the origin, i.e.,
$\mathbf{E}_\beta(-\mathbf{R})$.  Then this field can be taken outside the
integral and we just have
\begin{equation}
M_{\alpha,\beta}(\mathbf{R}) \sim \frac{\omega_{\text{p}}^2}{c^2}
\mathbf{E}_\beta(-\mathbf{R})\cdot\int \mathbf{E}_\alpha(\mathbf{x})d\mathbf{x},
\label{eq:mab4}
\end{equation}
where once again the integral runs over the cavity centered at the
origin.

Next, we attempt to calculate the relevant quantities needed to
solve for this matrix element.   In order to use the
tight-bind\-ing approach we will need to normalize the individual
eigenstates $\mathbf{E}_\alpha$.  Therefore, we will begin by obtaining
this normalization.  For $\ell = 1$, the $u_\ell(r)$'s are $r$ times
spherical Bessel functions.  We write this field as
\begin{equation}
E_{r,\text{in}} = +\frac{2C_1}{kr}j_1(kr)\cos\theta,
\end{equation}
where we have used the relation $P_1(\cos\theta) = \cos\theta$,
and introduced the normalization constant $C_1$, which will
be determined below.  Similarly,
\begin{equation}
E_{\theta,\text{in}} = -\frac{C_1}{kr}\frac{\partial[rj_1(kr)]}
{\partial r}\sin\theta,
\end{equation}
where we use $P_1^1(\cos\theta) = -\sin\theta$.   For $r > R$, we
have
\begin{equation}
E_{r,\text{out}} = +\frac{2D_1C_1}{k^\prime r}k_1(k^\prime r)\cos\theta
\label{eq:erout}
\end{equation}
and
\begin{equation}
E_{\theta,\text{out}} = -\frac{D_1C_1}{k^\prime r}\frac{\partial
[rk_1(k^\prime r)]}{\partial r}\sin\theta.
\label{eq:ethout}
\end{equation}

We will need the integrals of the {\em Cartesian} components of the
field over the volume of the sphere centered at the origin.  Let us
assume we are considering the $z$ mode, i.e., the one for which
$\theta$ refers to the angle from the $z$ axis.  Then the symmetry of
the problem shows that only the $z$ component of the electric field
will have a nonzero integral.  Also, we have that
\begin{equation}
E_{z,\text{in}}(r,\theta) = E_{r,\text{in}}\cos\theta
- E_{\theta,\text{in}}\sin\theta.
\end{equation}

Thus, after a little algebra, we find that the integral of this
field over the volume of the cavity is
\begin{equation}
\int E_{z,\text{in}}(r,\theta)d\mathbf{r} = \frac{8\pi C_1}{3k}R^2j_1(kR).
\end{equation}

Next, we work out the coefficient $D_1$. It is determined by the
boundary conditions at $r = R$.  These conditions are that $D_r$ 
and $E_\theta$ should be continuous at $r = R$.  These two
conditions determine not only the value of $D_1$ but also the allowed
frequency.  After a bit of algebra, we find that
\begin{equation}
D_1 = \frac{\omega^2}{c^2kk^\prime}\frac{j_1(kR)}{k_1(k^\prime R)} =
\frac{k}{k^\prime}\frac{j_1(kR)}{k_1(k^\prime R)}.
\end{equation}
The allowed value of $\omega$ is given by Eq.\ (\ref{eq:ws4}).

Finally, we need the normalization constant $C_1$.  We choose this so
that the integral of the square of the electric field for a
sin\-gle-cav\-i\-ty mode should be normalized to unity.  This condition
may be written
\begin{eqnarray}
& & \frac{2C_1^2}{k^2}\frac{4\pi}{3}\int_0^R\left[2j_1^2(kr) + \left(\frac{\partial
[rj_1(kr)]}{\partial r}\right)^2\right]dr \nonumber \\
& + & \frac{2C_1^2 D_1^2}{k^{\prime 2}}\frac{4\pi}{3}\int_R^\infty\left[2k_1^2(k^\prime
r) + \left(\frac{\partial[rk_1(k^\prime r)]}{\partial r}\right)^2\right]dr \nonumber \\
& = & 1.
\end{eqnarray}
If we write
\begin{equation}
\int_0^{kR}\left[2j_1^2(x) +
\left(\frac{d[xj_1(x)]}{dx}\right)^2\right]dx = F_1(kR)
\end{equation}
and
\begin{equation}
\int_{k^\prime R}^\infty\left[2k_1^2(x) +
\left(\frac{d[xk_1(x)]}{dx}\right)^2\right]dx = F_2(k^\prime R),
\end{equation}
then we can express the normalization condition as
\begin{equation}
\frac{8\pi C_1^2}{3}\left[\frac{F_1(kR)}{k^3} + D_1^2\frac{F_2(k^\prime
R)}{k^{\prime 3}}\right] = 1.
\end{equation}

Therefore, we can now write out an explicit expression for the matrix
element $M_{\alpha,\beta}(\mathbf{R})$ given in Eq.\ (\ref{eq:mab4}).  For the
$\alpha$th mode, the integral of $\mathbf{E}_\alpha$ over the volume
of a cavity is a vector in the $\alpha$th direction.  To evaluate
Eq.\ (\ref{eq:mab4}), we need the component of the $\alpha$th mode in the
$\beta$th direction at a position $\mathbf{R}$.  Let us first
consider the $z$th mode ($\alpha = z$).  We can use Eqs.\ (\ref{eq:erout})
and (\ref{eq:ethout}) to rewrite this field in Cartesian coordinates with the
additional equations
\begin{eqnarray}
\cos\theta & = & \frac{z}{r}, \nonumber \\
\sin\theta & = & \frac{\sqrt{x^2+y^2}}{r}, \nonumber \\
\hat{r} & = & \frac{x\hat{x}+y\hat{y}+z\hat{z}}{r}, \nonumber \\
\hat{\theta} &  = &\frac{xz\hat{x}+yz\hat{y}}{r\sqrt{x^2+y^2}} -
\frac{\sqrt{x^2+y^2}}{r}\hat{z}.
\end{eqnarray}
We just substitute these expressions back into Eqs.\ (\ref{eq:erout})
and (\ref{eq:ethout}) to get the Cartesian components of the field
for a mode parallel to the $z$ axis.  For the mode parallel to the $x$ axis,
we just permute the coordinates cyclically:  $z \rightarrow x$,
$x \rightarrow y$, and $y \rightarrow z$.   Similarly, for the $y$ modes,
we make the permutation $(x, y, z) \rightarrow (z, x, y)$.

Using these results, we should be able to compute all the elements
in the tight-bind\-ing matrix and hence obtain the band structure for
the photonic $p$-bands in the tight-bind\-ing approximation, in either
one or three dimensions.

\section{\label{sec:level3}Numerical Results}

\begin{figure}[ht!]
\begin{center}
\includegraphics[width=\columnwidth]{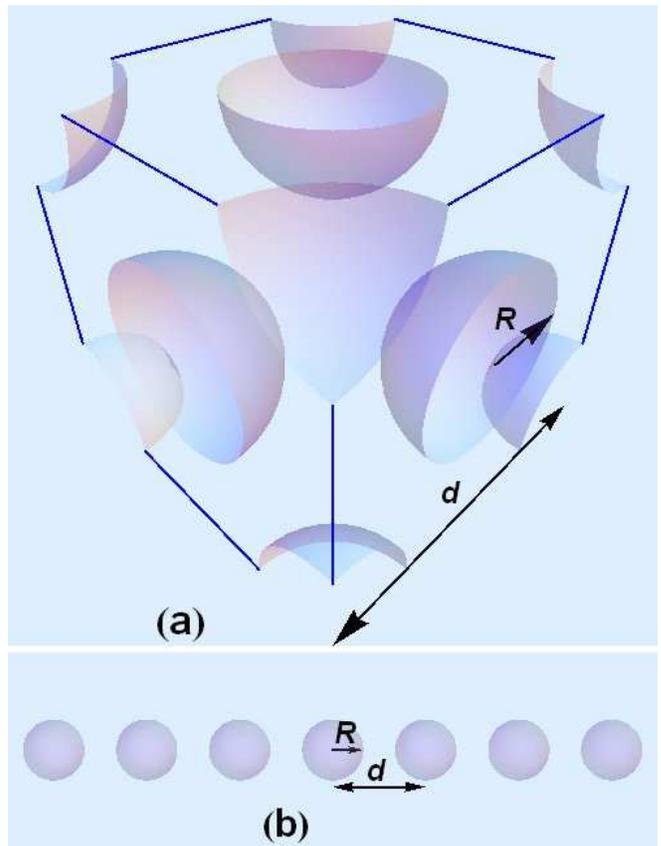}\\%
\end{center}
\caption{\label{fig:config}(Color online) Schematic diagram for (a) an inverse opal structure with a lattice constant $d$ and a void sphere radius $R$; (b)
a linear chain of nanopores with a pore separation $d$ and a nanopore radius $R$.}
\end{figure}

For the inverse opals we arbitrarily assume a lattice constant
$d=500\sqrt{2}\:\text{nm}$, and a void sphere radius
$R=150\:\text{nm}$ as in Fig.\ \ref{fig:config}(a).  This choice is the same as that of Ref.\ \cite{aliev}, 
where the Pb inverse opal has this lattice constant.  Since the volume of the primitive unit cell is
$v_{c}=d^{3}/4$, this corresponds to a void volume fraction $f =
0.160$. For the linear chain of nanopores (see below) this $d$ is the separation
between two nanopores and $R$ is the radius of a nanopore as in Fig.\ \ref{fig:config}(b).

Our band structures for the inverse opals are expressed in terms of the standard notation
for $\mathbf{k}$ values at symmetry points in the Brillouin zone. These
are $\Gamma=(0,0,0)$, $X=(2\pi/d)(0,0,1)$, $U=(2\pi/d)(1/4,1/4,1)$,
$L=(2\pi/d)(1/2,1/2,1/2)$, $W=(2\pi/d)(1/2,0,1)$, and
$K=(2\pi/d)(3/4,0,3/4)$.

The metallic dielectric functions we assume for the inverse opals and
linear chain of nanopores are of the usual Drude form,
\begin{equation}
\epsilon(\omega)=1-\frac{\omega_{\text{p}}^2}{\omega^2},
\label{eq:df}
\end{equation}
where $\omega_{\text{p}}$ is the plasma frequency of the
conduction electrons. $\epsilon(\omega)<0$ when
$\omega<\omega_{\text{p}}$, while $\epsilon(\omega)>0$ when
$\omega>\omega_{\text{p}}$.  Our calculations are thus carried out
assuming that the Drude relaxation time $\tau \rightarrow \infty$.
For a metal in its normal state, $\omega_{\text{p}}^2= 4\pi n
e^2/m$, where $n$ is the conduction electron density and $m$ is the
electron mass.  Note that with this choice of dielectric function,
the entire band structure can be expressed in scaled form.  That is,
the scaled frequency $\omega d/c$ is a function only of the scaled
wave vector $kd$, and the band structures are parameterized by the
two constants $\omega_{\text{p}}d/c$ and $f$ for the case of inverse opals.

Since we are considering void spheres in inverse opals and linear
chains of nanopores, it is of
interest to consider electromagnetic wave modes in a {\em single}
cavity, which could be considered a single ``atom'' of the void
lattice.
We show only results for $\omega < \omega_{\text{p}}$, since these
are the results most relevant to possible nar\-row-band photonic states
in the inverse opal structure.  Our  results for $\omega <
\omega_{\text{p}}$ for an isolated spherical cavity in an infinite
medium, and when $kR \ll 1$ and $k^\prime R \ll 1$ are given in
Table~\ref{tab:tm}.
These two inequalities are reasonable for our inverse
opal system parameters
$d=500\sqrt{2}\:\text{nm}$, $R=150\:\text{nm}$, and
$\omega_{\text{p}}d/c=1$, because
\begin{eqnarray}
kR & = & \frac{\omega}{c}R<\frac{\omega_{\text{p}}}{c}R
=\frac{\omega_{\text{p}}d}{c}\frac{R}{d}=\frac{3}{10\sqrt{2}}=0.2121, \nonumber \\
k^\prime R & = & \frac{\sqrt{\omega_{\text{p}}^2-\omega^2}}{c}R
=\sqrt{\left(\frac{\omega_{\text{p}}R}{c}\right)^2-\left(\frac{\omega R}{c}\right)^2} \nonumber \\
& = & \sqrt{\left(\frac{\omega_{\text{p}}d}{c}\frac{R}{d}\right)^2-(kR)^2}
= \sqrt{\left(\frac{3}{10\sqrt{2}}\right)^2-(kR)^2} \nonumber \\
& = & \sqrt{0.045-(kR)^2} < \sqrt{0.045}=0.2121.
\end{eqnarray}
The (modified) spherical Bessel functions in Eq.\ (\ref{eq:ws4})
are extremely close to the $\omega$ axis for $\ell > 5$,
so that
it is difficult to get eigenfrequencies for $\ell > 5$ in the isolated spherical
cavity.
However the eigenfrequencies continue to exist even for $\ell > 5$
when $kR \ll 1$ and $k^\prime R \ll 1$.

\begin{table}[ht!]
\begin{center}
\begin{tabular}{c|c|c} \hline \hline
\multicolumn{1}{c|}{}
 & \multicolumn{1}{c|}{Infinite medium}
 & \multicolumn{1}{c}{$kR\ll 1$, $k^\prime R \ll 1$} \\ \hline
$\ell=1$ & 0.1296 & 0.1299 \\
$\ell=2$ & 0.1232 & 0.1233 \\
$\ell=3$ & 0.1203 & 0.1203 \\
$\ell=4$ & 0.1186 & 0.1186 \\
$\ell=5$ & 0.1178 & 0.1175 \\ \hline \hline
\end{tabular}
\end{center}
\caption{\label{tab:tm}TM mode frequencies $\omega^\prime=\omega d/(2\pi c)$,
where $\omega < \omega_{\text{p}}$ and $\omega_{\text{p}}d/c=1$,
calculated for an isolated spherical cavity (``Infinite medium'') and those when
both $kR\ll 1$ and $k^\prime R \ll 1$. 
The (modified) spherical Bessel functions are extremely close to the $\omega^\prime$ axis for $\ell > 5$, so that
it is difficult to get eigenfrequencies for $\ell > 5$ in the isolated spherical
cavity. However this does not happen when $kR\ll 1$ and $k^\prime R \ll 1$.}
\end{table}
The solutions to Eq.\ (\ref{eq:wsteinfty}) do not exist for
$\omega < \omega_{\text{p}}$ with $\omega_{\text{p}}d/c=1$.
This fact is consistent with that the eigenvalues for $\omega < \omega_{\text{p}}$
do not exist for TE modes when $kR \ll 1$ and $k^\prime R \ll 1$.

\begin{figure}[ht!]
\begin{center}
\includegraphics[width=\columnwidth]{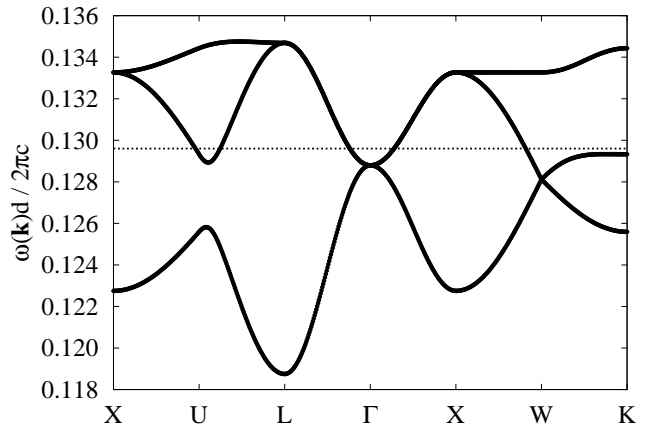}\\%
\end{center}
\caption{\label{fig:tb13_01}Tight-bind\-ing inverse opal band structure
for $\omega < \omega_{\text{p}}$
with $d=500\sqrt{2}\ \text{nm}$, $R=150\ \text{nm}$, and
$\omega_{\text{p}}d/c=1.0$, using $\omega_{\text{at}}d/(2\pi c)=0.1296$ for
$\ell=1$ in an infinite medium. The horizontal dotted line represents the
``atomic'' level.}
\end{figure}

For our fcc calculations, we calculate the
band structure including only the 12 nearest-neighbors of the cavity at
the origin.
Thus $\mathbf{R}=(d/2)(\pm 1,\pm 1, 0)$, $(d/2)(\pm 1,\mp 1, 0)$,
$(d/2)(\pm 1,0,\pm 1)$, $(d/2)(\pm 1,0,\mp 1)$, $(d/2)(0,\pm 1,
\pm 1)$, and $(d/2)(0,\pm 1,\mp 1)$. Assuming $\omega_{\text{p}}d/c=1$
and using $\omega_{\text{at}}d/(2\pi c)=0.1296$ for $\ell=1$ in
an infinite medium, we get the tight-bind\-ing results in Fig.\
\ref{fig:tb13_01}.  This figure shows three separate bands in the $X$-$U$-$L$ region
and $X$-$W$-$K$ region as expected for the $p$-bands. The bandwidth is
relatively small as $M_{\alpha,\beta}(\mathbf{R})d^2\sim 0.001$, which proves
the general relation between the bandwidth and the overlap integral.\cite{ashcroft}
All three bands are degenerate at $\mathbf{k}=0$ (the $\Gamma$ point).
In addition, there is a double degeneracy when $\mathbf{k}$ is directed
along either a cube axis ($\Gamma$-$X$) or a cube body diagonal ($\Gamma$-$L$),
the higher (concave upward) bands being degenerate in both cases.
The lower two bands have a band gap at the $U$ point, and these bands cross at the $W$ point.

\begin{figure}[ht!]
\begin{center}
\includegraphics[width=\columnwidth]{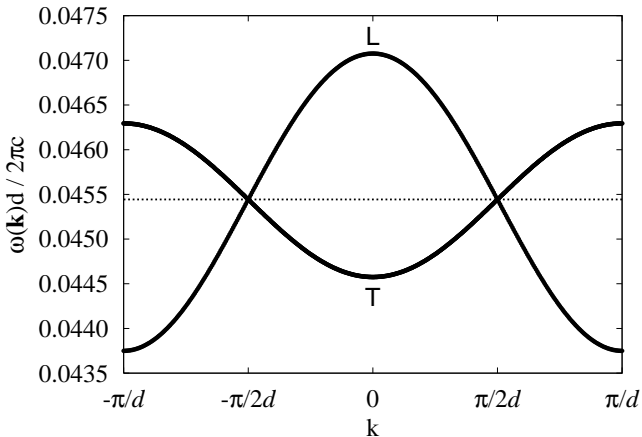}\\%
\end{center}
\caption{\label{fig:tb15_d}Tight-binding results of a periodic
chain of nanopores in a Drude metal host, for $\omega < \omega_{\text{p}}$.
We use $d=75\ \text{nm}$, $R=25\ \text{nm}$, and
$\omega_{\text{p}}d/c=0.35$, using $\omega_{\text{at}}d/(2\pi c)=0.0454$ for
$\ell=1$ in an infinite medium. Only the nearest-neighbors are included.
The horizontal dotted line represents the
``atomic'' level. In this and the following all plots, ``L'' and ``T'' denote
the longitudinal and transverse branches, respectively.}
\end{figure}

\begin{figure}[ht!]
\begin{center}
\includegraphics[width=\columnwidth]{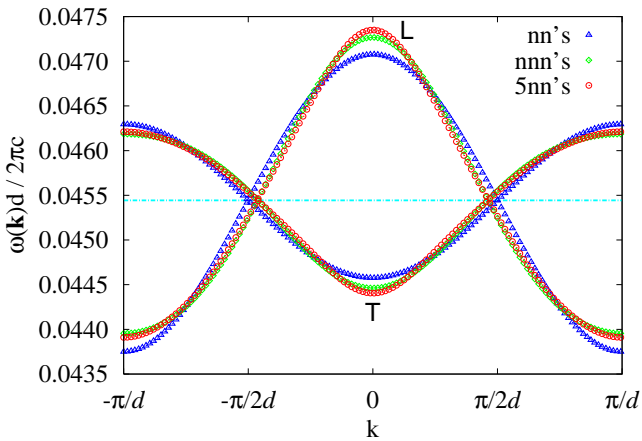}\\%
\end{center}
\caption{\label{fig:tb151719_01}(Color online) Same as Fig.\ \ref{fig:tb15_d}, but
including three different numbers of neighbors: nearest-neighbors (nn's),
next-nearest-neighbors (nnn's), and fifth-nearest-neighbors (5nn's).}
\end{figure}

Next, we turn to the band structure of a periodic linear chain of spherical nanopores in a Drude metal host.
For this linear chain, the Bravais lattice vectors are $\mathbf{R}=d(0,0,\pm n)$, where $\pm n$ is
the $n$th nearest-neighbor, $d$ is the separation between two nanopores and we
assume that the chain is directed along the $z$ axis.  We can calculate the
tight-binding band structure including as many sets of neighbors $\pm n$ as we wish.
To compare our results with those in Ref.\ \cite{brongersma}, we first use their
parameters, $R=25\ \text{nm}$ and $d=75\ \text{nm}$, together with their
overlap parameter $\omega_1=1.4\times 10^{15}\ \text{rad/s}$.  These combine to give
$\omega_{\text{p}}d/c=0.35$.  The ``atomic'' frequency is found by solving Eq.\ (\ref{eq:ws4}) and gives
$\omega_{\text{at}}d/(2\pi c) = 0.0454$ for $\ell = 1$ in an infinite medium.
Our resulting tight-binding dispersion
relations are shown in Fig.\ \ref{fig:tb15_d} with only nearest-neighbors included.
Note that our frequencies are given in unit of $2\pi c/d$ while the results
of Ref.\ \cite{brongersma} are not scaled.
Our results are exactly the inverse
images of theirs --- that is, we would get their curves (to within a constant of proportionality) if we reflect our curves 
through the horizontal line of the atomic level, 
and the transverse (T) branches 
are twofold degenerate as are theirs, while the longitudinal (L) branch 
is non-degenerate.
Our eigenfrequency for
a single-cavity $\omega_{\text{at}}$ corresponds to their resonance frequency $\omega_0$. As we increase the
number of nearest-neighbors (nn's) included, the separation between the $L$ and
$T$ branches increases at the zone center
but decreases at the zone boundary, as shown in Fig.\ \ref{fig:tb151719_01};
the same trend is seen in Fig.\ 1 in Ref.\ \cite{brongersma}.
The sum also converges quickly, so there is
little difference between the dispersion relation
including through the next-nearest-neighbors and that including through the 5th nearest-neighbors.



\begin{figure}[ht!]
\begin{center}
\includegraphics[width=\columnwidth]{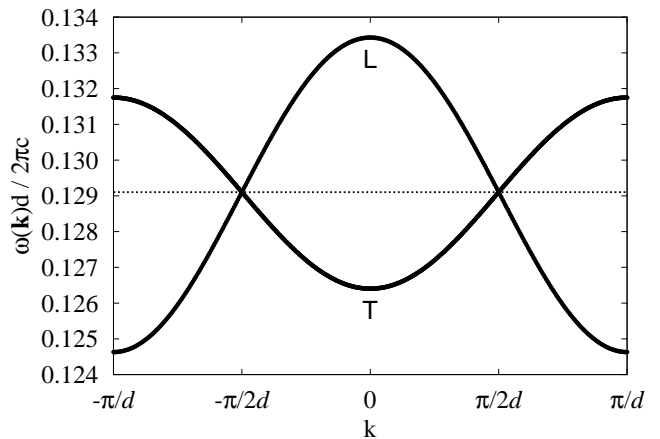}\\%
\end{center}
\caption{\label{fig:tb16_01}Same as Fig.\ \ref{fig:tb15_d},
except $\omega_{\text{p}}d/c=1.0$ and $\omega_{\text{at}}d/(2\pi c)=0.1291$.}
\end{figure}

\begin{figure}[ht!]
\begin{center}
\includegraphics[width=\columnwidth]{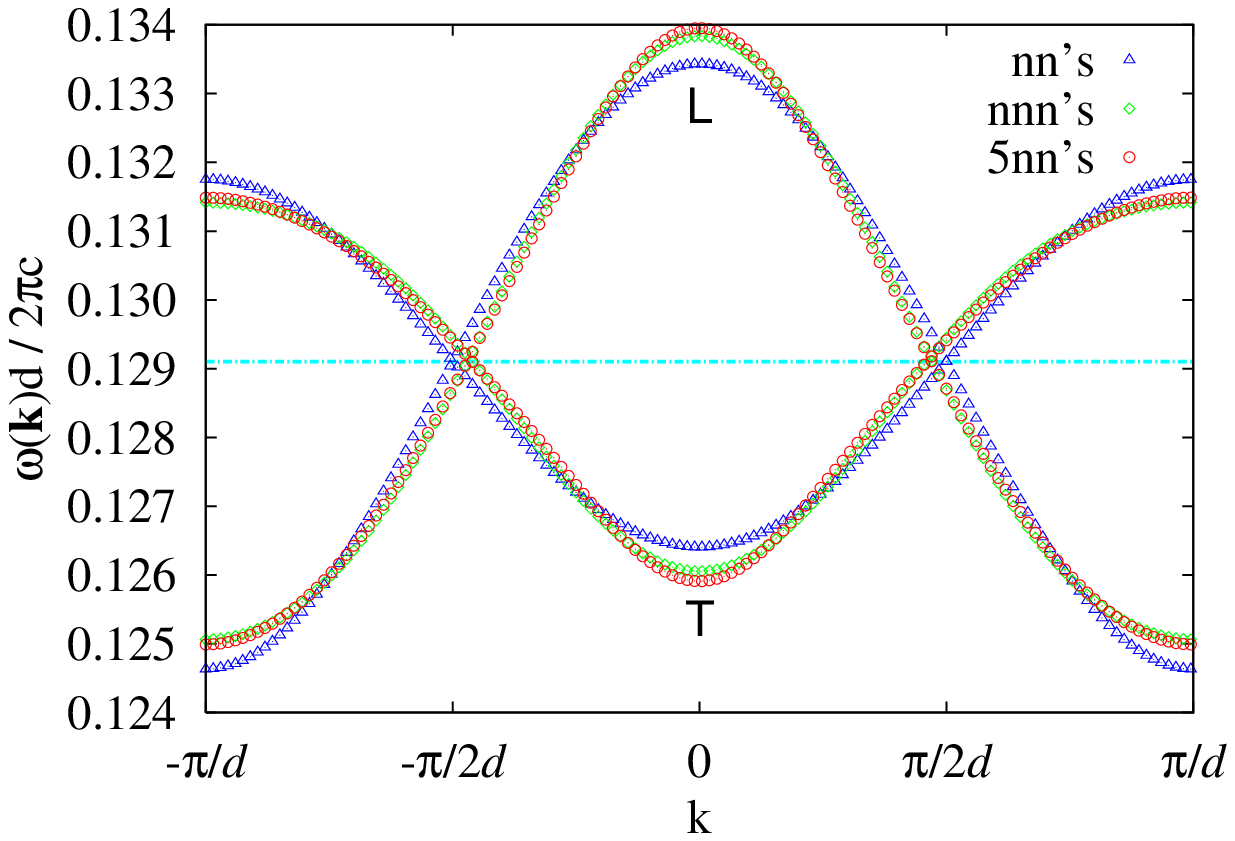}\\%
\end{center}
\caption{\label{fig:tb161820_01}(Color online) Same as Fig.\ \ref{fig:tb151719_01},
except $\omega_{\text{p}}d/c=1.0$ and $\omega_{\text{at}}d/(2\pi c)=0.1291$.}
\end{figure}

\begin{figure}[ht!]
\begin{center}
\includegraphics[width=\columnwidth]{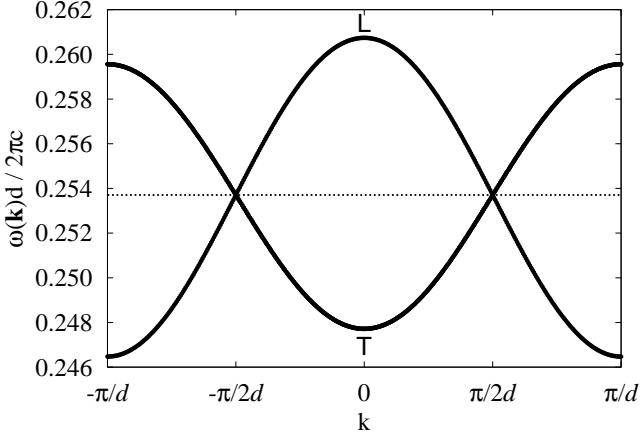}\\%
\end{center}
\caption{\label{fig:tb21_01}Same as Fig.\ \ref{fig:tb15_d},
except $\omega_{\text{p}}d/c=2.0$ and $\omega_{\text{at}}d/(2\pi c)=0.2537$.}
\end{figure}

\begin{figure}[ht!]
\begin{center}
\includegraphics[width=\columnwidth]{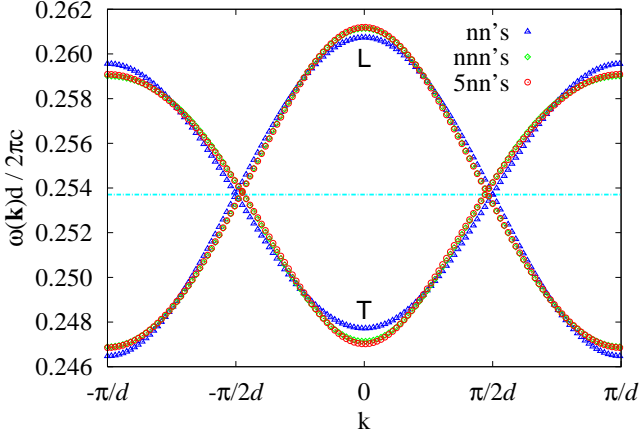}\\%
\end{center}
\caption{\label{fig:tb212223_01}(Color online) Same as Fig.\ \ref{fig:tb151719_01},
except $\omega_{\text{p}}d/c=2.0$ and $\omega_{\text{at}}d/(2\pi c)=0.2537$.}
\end{figure}

\begin{figure}[ht!]
\begin{center}
\includegraphics[width=\columnwidth]{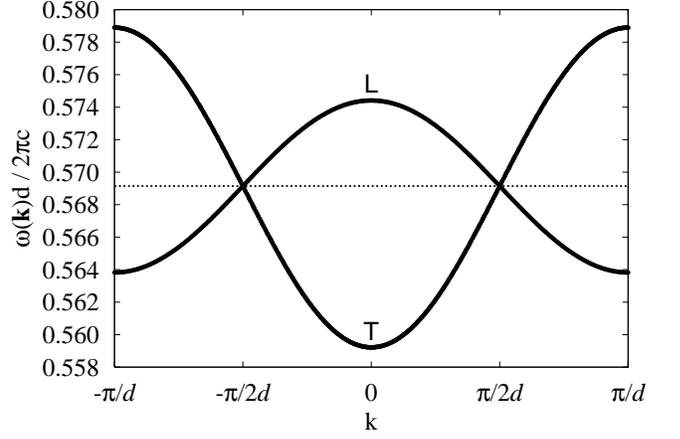}\\%
\end{center}
\caption{\label{fig:tb24_01}Same as Fig.\ \ref{fig:tb15_d},
except $\omega_{\text{p}}d/c=5.0$ and $\omega_{\text{at}}d/(2\pi c)=0.5691$.}
\end{figure}

\begin{figure}[ht!]
\begin{center}
\includegraphics[width=\columnwidth]{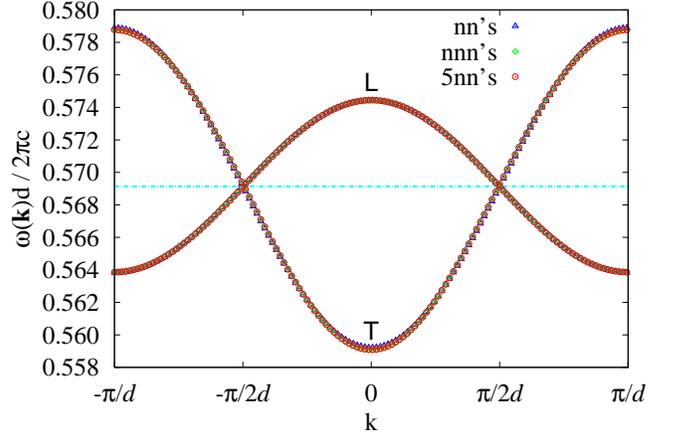}\\%
\end{center}
\caption{\label{fig:tb242526_01}(Color online) Same as Fig.\ \ref{fig:tb151719_01},
except $\omega_{\text{p}}d/c=5.0$ and $\omega_{\text{at}}d/(2\pi c)=0.5691$.}
\end{figure}

We have carried out similar calculations using other values of
the parameter $\omega_{\text{p}}d/c$, namely $1.0$, $2.0$, and $5.0$.
Such calculations are possible here because our calculations are
non-quasistatic, so
that the overlap integral between neighboring spheres falls off exponentially
with separation.
The results are given in Figs.\ \ref{fig:tb16_01}, \ref{fig:tb21_01}, and \ref{fig:tb24_01}, 
respectively. The corresponding results including more overlap
integrals are shown in Figs.\ \ref{fig:tb161820_01}, \ref{fig:tb212223_01}, and \ref{fig:tb242526_01}, respectively.
It is also striking that, as $\omega_{\text{p}}d/c$ increases in going from Fig.\ \ref{fig:tb15_d} to 
Figs.\ \ref{fig:tb16_01}, \ref{fig:tb21_01}, and \ref{fig:tb24_01},
the ratio $r_{\text{LT}}$ of the width of the L band to that of the T band steadily decreases.  In 
Fig.\ \ref{fig:tb15_d}, $r_{\text{LT}} > 1$, in Fig.\ \ref{fig:tb24_01}, $r_{\text{LT}} < 1$, while in 
Fig.\ \ref{fig:tb21_01} (for which $\omega_{\text{p}}d/c = 2.0$), $r_{\text{LT}} \sim 1$.   

One could also say that, except for an overall scale factor, Fig.\ \ref{fig:tb24_01}
looks like an inverted image of Fig.\ \ref{fig:tb15_d} about the horizontal line of $\omega_{\text{at}}$.
The dispersion relations for the intermediate value $\omega_{\text{p}}d/c=2.0$ has nearly perfect symmetry about the horizontal line of $\omega_{\text{at}}$ (the T
branches are nearly reflections of the L branch about the horizontal line of $\omega_{\text{at}}$), as in
Fig.\ \ref{fig:tb21_01}.
For the nn case, the T and L bands cross at $\pm \pi/(2d)$, as can be seen in Figs.\ \ref{fig:tb15_d}, \ref{fig:tb16_01}, \ref{fig:tb21_01}, and \ref{fig:tb24_01}.  When further neighbors are included, they cross at smaller values than $|\pi/(2d)|$, as can be seen in Figs.\ \ref{fig:tb161820_01}, \ref{fig:tb212223_01}, and \ref{fig:tb242526_01}, but the crossing points get closer to $\pm \pi/(2d)$ as $\omega_{\text{p}}$ increases.  Also, the effects
of including further neighbors become smaller as $\omega_{\text{p}}$ increases; 
they are smallest at $\omega_{\text{p}}d/c=5.0$, as can be seen in Fig.\ \ref{fig:tb242526_01}.

%
%
%
%
%
%
%
%
%
%
%
%


\begin{figure}[ht!]
\begin{center}
\includegraphics[width=\columnwidth]{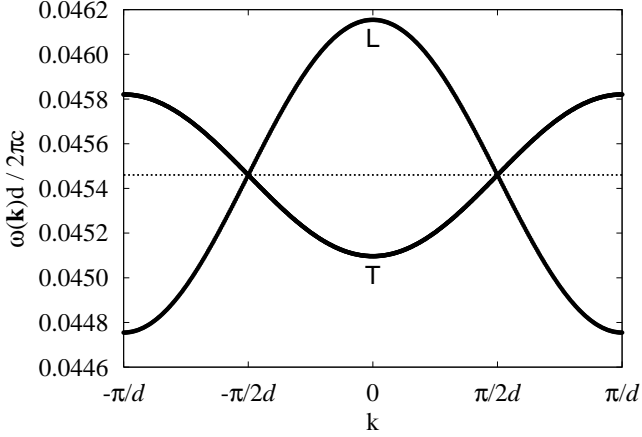}\\%
\end{center}
\caption{\label{fig:tb15_01_d}Same as Fig.\ \ref{fig:tb15_d},
except that $R/d=0.25$ and $\omega_{\text{at}}d/(2\pi c)=0.04546$.}
\end{figure}

\begin{figure}[ht!]
\begin{center}
\includegraphics[width=\columnwidth]{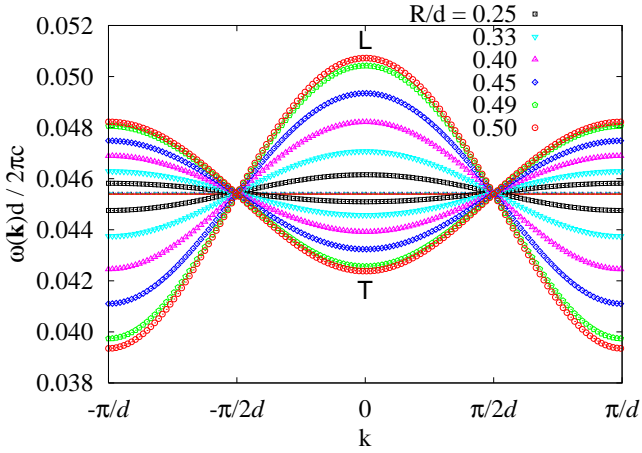}\\%
\end{center}
\caption{\label{fig:tb1512345_01}(Color online) Plotting together six different results for $\omega < \omega_{\text{p}}$, all
with $\omega_{\text{p}}d/c=0.35$, but with different $R/d$:
$R/d=0.33$ and $\omega_{\text{at}}d/(2\pi c)=0.04544$; $R/d=0.40$ and $\omega_{\text{at}}d/(2\pi c)=0.045426$;
$R/d=0.45$ and $\omega_{\text{at}}d/(2\pi c)=0.045412$; $R/d=0.49$ and $\omega_{\text{at}}d/(2\pi c)=0.045399$;
and $R/d=0.50$ and $\omega_{\text{at}}d/(2\pi c)=0.045396$. In each case, only nearest-neighbor overlaps are included.}
\end{figure}

\begin{figure}[ht!]
\begin{center}
\includegraphics[width=\columnwidth]{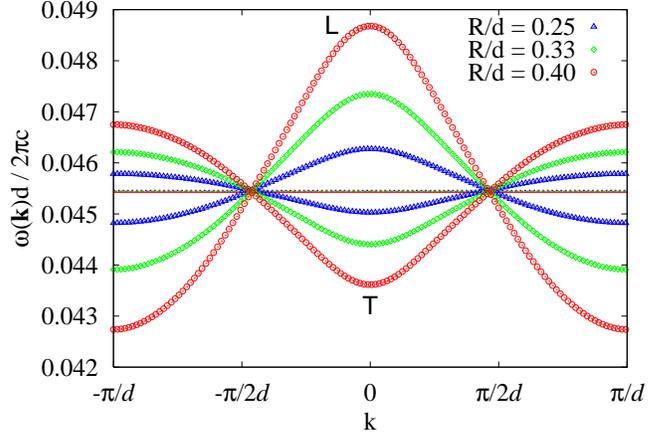}\\%
\end{center}
\caption{\label{fig:tb19012_01}(Color online) Same as Fig.\ \ref{fig:tb1512345_01},
except that only three $R/d$'s are plotted, with inclusion of up to the fifth
nearest-neighbors.  We omit the three largest values of
$R/d$ because it may be necessary to include more than just $\ell = 1$ when
$R/d \gtrsim 0.4$.}
\end{figure}

Next we consider values of $R/d$ other than $1/3$, but still keeping the same
value of $\omega_1=1.4\times 10^{15}\ \text{rad/s}$ (i.e., $\omega_{\text{p}}d/c=0.35$).
For a smaller $R/d=0.25$, the variation of the band energies with $k$
becomes smaller, as seen in Fig.\ \ref{fig:tb15_01_d}, than it is in
Fig.\ \ref{fig:tb15_d}, but the crossing points between the L and T branches still occur at $\pm \pi/(2d)$. This behavior
becomes clearer when the results for several values of $R/d$ are plotted together as in Fig.\ \ref{fig:tb1512345_01}. As
$R/d$ increases, the variation of the band energies with $k$, and the separation  between the L and T branches at both the zone center and zone boundary, increase,
but the L and T branches still cross at $\pm \pi/(2d)$. If we include more neighbors up to fifth nearest-neighbors, but consider only
up to $R/d = 0.4$, we get the dispersion relations shown in Fig.\ \ref{fig:tb19012_01}.  These show the same trends as in
Fig.\ \ref{fig:tb1512345_01}, except that the band crossing points occur at values of $|k|$ slightly less than $|\pi/(2d)|$.  Furthermore, the
separation between the L and T bands increases slightly at $k = 0$, but decreases
slightly at $k = \pm \pi/d$.  We show only $R/d$ up to $0.4$ in this
Figure because, in the quasistatic limit, there is evidence that for larger
values of $R/d$ the dispersion relations are significantly modified by higher
values of $\ell$.\cite{park}

\section{\label{sec:level4}Discussion}

In this work we have calculated the photonic band structures of
metal inverse opals and of a linear chain of spherical voids in a metallic host
for frequencies below $\omega_{\text{p}}$, when $\ell=1$ using a tight-bind\-ing approximation.
In both cases, we include only the $\ell = 1$ ``atomic'' states of the voids.
As a possible point of comparison, we have also computed the same band structures
using the asymptotic forms of the spherical and modified spherical Bessel
functions for small void radius.
In this asymptotic region, there are only TM modes.
The results for the linear chain of voids can be considered as the
``inversions'' of those in Ref.\ \cite{brongersma}, in the sense discussed
earlier. In other words, if we reflect our L and T branches with respect to
the atomic energy level, we would get their L and T modes.


Although we did not discuss this approach, we did attempt to use the plane
wave expansion method to calculate the band structure for the inverse opals,
similarly to Refs.\ \cite{mcgurn} and \cite{kuzmiak1}.  Just as found
in those papers, the photonic bands for modes below
$\omega_{\text{p}}$ depend on the number of plane waves included in
the expansion and on the type of field, $\mathbf{B}$ or $\mathbf{E}$,
used in the expansion.
Furthermore, this plane wave expansion method gives a large number of
flat bands below $\omega_{\text{p}}$, which are difficult to interpret physically.
Because of this problem, and because of the apparent non-convergence of this
approach with the number of plane waves, we do not present these results here.
By contrast, the band structures above $\omega_{\text{p}}$, when calculated
using the plane wave expansion, varied
smoothly with $\mathbf{k}$ and converged well with the number of plane waves
included.


In calculating the tight-binding band structure for the linear chain (and the inverse opal structure), one should, in principle, include
all the neighbors.  But in practice, for the linear chain, it is sufficient
to include only up to the fifth nearest-neighbors.  This calculation is easily
carried out, since the matrix in Eq.\ (\ref{eq:mab}) is already diagonal in $x$, $y$, $z$ and the sum converges quickly. In fact, even the inclusion of neighbors beyond the first two sets changes the band structure very little.  The smallness of
the further neighbor effect is particularly apparent when $\omega_{\text{p}}d/c=5.0$, as in Fig.\ \ref{fig:tb242526_01}.

Although we studied the 3D and 1D lattices, we have not investigated a 2D lattice
of spherical pores in a metallic host.
For the 2D case, the matrix element $M_{\alpha,\beta}(\mathbf{R})$ can again be readily calculated using our tight-binding approximation. It is expected to have some nonzero off-diagonal elements in addition to the
diagonal elements.  Thus band structures somewhat similar to the 3D band structures shown in Fig.\ \ref{fig:tb13_01} are also expected in the 2D case.

In the quasistatic case, for metal grains in air, when $R/d$ is greater than about $0.4$,
it becomes important to include more than just $\ell = 1$ as in Ref.\ \cite{park}.
Inclusion of such higher $\ell$'s might be rather difficult
in the present dynamical case, though it would be possible in the quasistatic
limit for 1D chains of spherical nanopores.

In the present work, we have considered only the case of
one cavity per primitive cell and $\ell = 1$.  It
would be of a great interest to consider multiple cavities per unit cell.
Of course, in this case, the dimension of the matrix in Eq.\ (\ref{eq:mab}) will increase and
the Bloch states in Eq.\ (\ref{eq:nbs}) will acquire an additional index.
It should be straightforward to extend the present work to such a case,
which would make an interesting subject for future work.

The surface effect of the metal nanopores on the eigenstates becomes prominent as
the radius of a pore ($R$) or the ratio of radius to center-to-center separation
($R/d$) increases. Shockley surface states form on metal surfaces depending on the
type of metal, such as the Fermi wavevector $k_{\text{F}}$ or the Fermi energy
$\epsilon_{\text{F}}$. These surface states will interact with the eigenstates,
resulting in the change of eigenvalues. However we think these effects are negligible,
especially in the quasistatic limit since in the asymptotic limit, these oscillatory
interactions take the form\cite{hyldgaard}
\begin{equation}
E_{\text{pair}}^{\text{asym}}(a) \propto \left(\frac{4\epsilon_{\text{F}}}
{\pi^2}\right)\frac{\sin(2k_{\text{F}}a+2\Theta)}{(k_{\text{F}}a)^2},
\label{eq:epair}
\end{equation}
where $a$ is the interatomic separation, $\Theta$ is the effective
interaction phase shift, $k_{\text{F}}$ the Fermi wavevector of the isotropic
surface state, and $\epsilon_{\text{F}}$ the Fermi energy. The proportionality constant gives the consequences of scattering into bulk states.\cite{hyldgaard} The very slow $a^{-2}$ decay of these interactions allows them to play a role at large separations, but the overall magnitude is small due to the sinusoidal term.

In summary, we have described a tight-binding method for calculating the photonic band structure of a periodic composite of spherical pores in a metallic host, and
have applied it to both 1D and 3D systems.  The method is fully dynamical, and is not limited to very small pores.  The method does not have the convergence problems found when the magnetic or electric field is expanded in plane waves.  Furthermore, there are no radiation losses to consider, unlike the complementary case of small
metal particles in an insulating host, because the fields associated with these
modes outside the pores are exponentially decaying.  Thus, this method may be useful for a variety of periodic metal-insulator composites.  It would be of interest to compare these calculations to experiments on such materials.

\begin{acknowledgments}
This work was supported by NSF Grant No.\ DMR04-13395 and by an NSF MRSEC grant at the Ohio State University,
Grant No.\ DMR08-20414. 
All of the calculations using plane wave expansions
were carried out on the P4 Cluster at the Ohio Supercomputer Center, with
the help of a grant of time.
\end{acknowledgments}

\end{document}